# Examination of the scientific review process: Ten "best practice" suggestions for an improved process

Dr. Ilana Harrus[1]

**Abstract:**

   In this article we wish to provide a common set of best practice approaches that should be considered for all effective research grant proposal reviews.  The federal government performs a critical role in American competitiveness and security by supporting basic research funded with taxpayer dollars. Effectively managing their allocation to scientists and researchers is a noble and crucial mission for advancing fundamental knowledge and deserves a heightened attention. Ensuring that proposals submitted are treated fairly and transparently is essential to both the health of any research program and also a duty to the public who ultimately funds the research.
      **The paper describes the general requirements of a review process and at each step underlines the issues and suggests potential improvements and some fundamental requirements that should be included in any scientific review. We also included a series of tips geared to the scientific community. Our goals in this paper are 1) to demystify the process for everyone including policy makers who are sometimes flummoxed by the results of some scientific reviews, 2) to trigger some discussions about reviews and review process in the scientific community, 3) to inform scientists whose careers are directly impacted by review results about their own role in this process and 4) to suggest a road to a more efficient, fairer and overall more transparent process.**
For experts in proposal reviews or for busy or impatient readers, the entire list of our recommendations is presented at the beginning.  We describe in each section the context and rational of each recommendation.

 Dr. Harrus is an astrophysicist who has been involved in proposal reviews in various capacities for about 10 years including serving as a program scientist in the Astrophysics Division of the National Aeronautics and Space Administration (NASA) Headquarters and as a program director at the National Science Foundation (NSF) both in the Astronomy and in the Physics divisions.  In her career, she has participated in more than one hundred reviews taking on all possible roles from proposer to reviewer to panel chair and from organizer of a single panel to oversight of an entire program review.

---

[1] The author can be contacted at: ilana.harrus@gmail.com

# List of Best Practice Suggestions (BPS).

**BPS1: A proposal should be assigned to a panel at least three weeks before the review. The assignment time should be part of the public report on that proposal.**

**BPS2: Adopt a system of self-grading the non-conflicted reviewers on each of the proposals assigned to the panel. The average for each reviewer and average for each proposal <u>should be kept as part of the official review record</u> to prove, should contestation arise afterwards, that all proposals received a fair and equal treatment and that the panel was indeed adequate.**

**BPS3: Keep the identity of the panelists completely anonymous until the first day of the review. This includes preventing them from reading the reviews submitted by other panelists. This completely eliminates the danger of "pre-review" talks or influence between panelists without the oversight of a review official.**

**BPS4: Any Program Director ruled in conflict with a proposal should have absolutely nothing to do with that proposal. This includes assigning reviewers to that proposal.**

**BPS5: The number of people participating <u>in the final discussion and grading</u> on a proposal should be tracked and monitored across programs and panels. That number should be made part of the record available on each proposal should a contestation be received after the review.**

**BPS6: The grading of a proposal should be numerical. The grade should be the average of the grades given by the non-conflicted members of the review panel. The vote should be secret (grades written on a piece of paper and passed to the agency representative). The average obtained for each proposal should be kept secret until the end of the first discussion session.** *The range is left at the discretion of the program but we recommend nothing smaller than a 1 to 5 range and using the entire range with at least one decimal place.*

**BPS7: The final list of proposals assigned to each panel should be made public (or at least available upon request) with all information (the entire line) about the proposals not recommended for funding blanked out. Only the proposals pushed forward should appear on the list in the order the panel recommended them and with their overall ranking or grade. Even if most of the list is redacted, it will show the final ranking of funded proposals.**

**BPS8: Implement a unified method for the final ranking of the panel. We suggest a Modified Borda Count (MBC) method because it is a proven method combining the advantages of taking care of conflict of interest, ranking and consensus all in one step.**

**BPS9: All recommendations for funding to the selecting official should be co-signed by more than one program director. We recommend at least three with one not directly connected to the program but knowledgeable about the science. The written document should be made public with confidential information redacted when a protest or inquiry arises.**

**BPS10: Implement a performance measurement to include a short internal limit of about 90 days to organize a review. Add two additional longer time ranges to review most and then all proposals. Reviews should be done and closed in at most 6 months.**

# **Introduction**

All proposal reviews have the same declared goal: Help the funding agency decide which proposals among the thousands submitted each year are worthy of public funding. Past this basic commonality, pretty much everything else varies not only between federal agencies but also within an agency at all levels, sometimes even within a specific program.

In fact the subject of review process is so filled with "it depends" that an entire book would be required to fully understand the mechanics of the different reviews. In this article we wish to provide a "basis", a common set of prescriptions that we argue should be included in all reviews. Most of these prescriptions have been tested and proven in actual reviews. Some may seem trivial or too "detailed" but the proposed common set of best practice approaches should be present in all reviews to achieve a Research and Development (R&D) portfolio allocation of the highest quality. In some cases we have ignored some fine distinctions that are beyond the scope of this article or have restricted ourselves to particular examples. We hope that the experts reading this article will understand the need for the "shortcuts" we sometimes took in our search for concision. In some other cases we do go into more details especially when these are important to the process or can provide some insight to reviewers.

The paper is organized practically as the chronology of a review, inserting when necessary our recommendations after describing the reason for that recommendation. We argue that these should constitute the basis of any respectable review.

We wish to provide the Principal Investigators (PI)s, the lead researchers in research grant proposals, a clear understanding of the reviews they are facing. When necessary we have inserted in the text some "tips" for the PIs. These appear in small separated shaded "boxes". We also hope that reviewers will look at their task in a more informed way. Finally we hope that people in charge of the process will assess more rigorously their responsibilities toward the community. It is our belief that even if "there is no perfect process" this often-used sentence should not be an excuse to repeat the same mistakes years after years. There may never be a perfect process but that should not prevent us for trying to improve it every day.

# **1) Proposal review 101**
## **1-1 Parties involved in a review.**

Any well-organized proposal review should involve at least three different groups: A group of people organizing the review, a group of people participating in the review as reviewers and a group of people submitting their proposals for review. As obvious as it may seem, even that very simple premise is not always true. At NSF for example, there are proposals that can be funded without any internal or external review and consent beyond the decision of a single person, most commonly a program director[1].

---

[1] NSF Grant Proposal Guide (2014), Pages II-23:II-29

Another rule that may seem obvious to the external observer but which is sometimes also violated, stipulates that these groups be completely disjoint. In other words, no one should be at the same time a reviewer and a proposer or an organizer and a proposer. The fact of being an organizer and a reviewer is not seen as dramatically impacting the review except of course if this person is also a proposer! Some reviews allow people to participate even if they have a proposal in competition. This is in fact the basis of a pilot review process based on game theory and implemented at NSF[1], and there are even reviews where people organizing the review, either at the funding agency or as part of the mission or telescope allocating the time and/or the money, can be named in proposals for review.

In the remaining portion of the article, we will move away from such overlap[2] and assume that the review has these three groups in place and that they are indeed disjoint. We will first briefly present the three groups and then go into the details of interactions between all of them and their roles.

1-1-1 <u>The organizers.</u>

The first is the group of people in charge of the review. This can include members of the funding agency or representatives of a mission or a program. At the very top of this group is a selecting official whose signature is on the official documents engaging the funding agency. Contrary to a widespread perception, participants to a review panel **<u>do not</u>** accept or reject proposals. That prerogative belongs solely to the "Selecting Official" who signs the documents engaging the funding agency's responsibility. For example, at NSF, the selecting official for most grant programs can be either the Director or the Deputy Director of the Division of Astronomical Sciences[3]. They review the suggestions of their program directors, or program scientists in NASA parlance[4], and concur or not on their recommendations. Selecting officials are seldom engaged in the complete details of the review process as they may have none or very little formal experience in reviews and review process, but they do have an oversight function. The depth of their involvement usually reflects more their interest in a particular review (not all reviews are created equal) and/or their management style. The other players in this first group are the program directors, or program scientists, whose job it is to organize the review according to the guidelines of their agency and after completion of the review, debrief the selecting official with their recommendations.

Programs directors are de-facto the link between the community and the funding. They organize the reviews, sorting proposals by panels and deciding on the dates, location and format of the reviews. They recruit the reviewers, attend the reviews and in some cases orchestrate the debates. They also formulate and write the arguments for or against a proposal. In short, their role is crucial to the review process. They can be civil servants or employed temporarily (from 2 to 4 years at NSF, 2 to 6 years at NASA) under an Intergovernmental Personnel Act (IPA) agreement with their home institution. We will examine every single one of their tasks in details later but first let's review briefly the other two groups involved in a review.

---

[1] Report to the National Science Board on NSF's Merit Review Process Fiscal Year 2013 – Page 46 (Mechanism Design).
[2] We kept away from these for clarity and concision but hope that some others will pay close attention to these reviews.
[3] In the case of major programs both NSF director and MPS director can get involved or actually **be** the selecting official.
[4] In this article, we are using both terms interchangeably although we will point, when needed, the major differences between the two roles.

1-1-2 Reviewers

Reviewers are usually recruited by the program director in charge. In theory, reviewers are the experts of the subjects covered by the proposals in that particular panel. In practice the most stringent constraints on panel members are not their expertise but both their availability and willingness to participate in the review and the potential conflict of interest with proposals in their panel or in the review. This is true to such an extent that the practice of recruiting reviewers before proposals are even submitted is quite widespread. People recruited this way tend to be "generalists" because no one should recruit an expert in a particular subject without knowing if proposals on that subject are indeed submitted. This is done with the assumption that a panel could be "fine-tuned" when the information about conflicts of interest becomes available and that a generalist "in the field" is as useful for a review as an expert.

> **Tip #1: (for the young and untenured crowd)**
>
> **Before** you accept to be part of a review, ask the program officer to provide you with an official letter saying that you will be participating in the proposal review for the named program. The letter should **not** specify the actual panel, nor the date of the review, only the name and the year of the program. Make it clear that you are not asking for an "endorsement letter" which is illegal to provide, but a personalized proof that you are being tasked by the funding agency. Keep these letters.

Reviewers are expected to do an incredible amount of work. Most program directors ask them to read all the proposals in their panel, grade and write a report on the subset of proposals assigned to them, and be prepared to argue in front of colleagues the merits or problems of each proposal. This is in addition to the time they have to spend to participate, physically or remotely, to the actual panel review. It is also worth noting that reviewers are usually either paid a nominal fee, reimbursed their expenses or simply no paid for their work. It is then no surprise that it takes time to recruit competent people for reviews and that the frustration of not being able to convince someone to participate can sometimes boil over. I recall a young IPA at NSF threatening privately to retaliate against a scientist he failed to recruit for an upcoming review even though, and this seemed to be the source of much of his surprise at her reluctance to accept on the spot, she still had a proposal under consideration in another of the programs he was administering.

Possibility of blackmail put aside, it is a good idea for scientists to participate in reviews. The first and by far the most important reason is that panel reviews provide a good overview of what other people are proposing, both the good and the bad, and a lesson in the mechanism of the process. Second it gives reviewers some opportunity to network. This is less true at NSF than at NASA because reviews at NSF are usually done either remotely (in FY2013, 28% of all reviews were done remotely[1]) or one panel at a time so networking opportunities are considerably reduced compared to NASA where a review can have anywhere from 30 to 80 participants all gathered in one place for a couple of days[2]. Third, it does give the reviewer an opportunity to meet and interact with their discipline's program director and others at the funding agency. We will go back to the details of the reviewer's work and the important differences existing between reviewers according to the specific reviews after briefly mentioning the last (but certainly not the least) group in any review process.

1-1-3 PIs and Cie.

The lead person responsible for the scientific or technical completion of a proposal project is called a Principal Investigator (PI). PIs are of course the most important piece of this process. PIs though are treated very differently by the different funding agencies. Before examining the details of their difference, let's briefly outline the general similarities of PIs across agencies. Most PIs submit proposals in response to a "call for proposals"[3]. There are many different types of calls and a complete discussion on the commonalities and differences of the different calls is outside the scope of this paper. Suffice to say that each call usually includes a guideline on who does or does not qualify to submit a proposal. Setting aside the issues of the PI's affiliation, as grants are issued to the PI's institution and not the PI itself, NASA Astrophysics' rules (for example) exclude most proposals and PIs not connected in some ways to NASA mission or science. This ensures that all proposals reviewed reflect NASA's and NASA's Astrophysics priorities which are defined and refined continuously. In contrast, the general grant program (called AAG) has an explicit statement allowing all astronomy-based proposals even those not included in the four main categories of research areas underlined by NSF. This inclusiveness is rooted in the creation charter of NSF directing the Foundation "to initiate and support basic scientific research and programs to strengthen scientific research potential and science education programs at all levels"[4].

This means that there is no strict restriction on proposals submitted to NSF AAG program[5]. We have until now only mentioned the PIs for any proposal submitted. In fact, proposals submitted by only one person are relatively rare and most involve co-PIs, co-Is (for NASA proposals, NSF does not have co-I per-se), collaborators and many people not always named in the proposal.

---

[1] Reference: Report to the National Science Board on the NSF's Merit Review Process Fiscal Year 2013. Page 45.
[2] Some NASA reviews are also done remotely but less frequently than at NSF.
[3] There are exceptions to this but a detailed description of the proper handling of "unsolicited proposals" is outside the scope of this article.
[4] Reference: "National Science Foundation Act of 1950" or Public Law 507 (81st Congress)
[5] This could change very soon. With the recent decreasing acceptance rates, both agencies are exploring ways to limit the number of proposals submitted (and reviewed) - See for example: Dr. Ulvestad – Public Presentation to the AAAC (June 2014.), http://www.sciencemag.org/content/344/6190/1328.full or more recently:http://arxiv.org/abs/1510.01647)

Most proposals are headed by a PI/PD (Principal Investigator or Project Director). This person is the main individual responsible for the scientific or technical completion of the project. The co-PI shares this responsibility. We note that a co-PI in the NSF parlance is different from a co-I in the NASA vocabulary. A co-I for a NASA proposal is someone that will contribute to the project and will receive money for his/her work. A co-I could be based at a different institution than the PI and the money would be set as a sub-award from the PI's institution to that of the co-I.

A co-PI, in contrast, shares with the PI the entire responsibility of the proposal. The PI in that case is simply the designated point of contact for the communications from NSF to the project (NASA does not have official co-PIs but several awards are sometimes made directly instead of having sub-awards going through one single institution). Some NSF proposals are collaborative proposals with one project PI, several co-proposals PIs and all of the co-PIs associated with each proposal. Many of the other people involved in a proposal, that is to say, people who are getting paid by the project, can be included in the budget only and not in the cover page that populates the database of proposal submissions.

> **Tip #2:**
>
> PIs should be very careful when considering the PI/co-PI designation because co-PIs have all the responsibilities of a PI. In addition, at NSF co-PIs are excluded from the entire review while a (paid) collaborator can participate (except of course for that particular proposal or panel).

**1-2 Mechanics of the review**

1-2-1 Timeline of a review

All proposal reviews, however different they are, share common features. One of these is what needs to be done right after the deadline is passed[1]. As soon as the proposal is received, usually taken as the deadline date for all proposals, a clock starts "ticking". For both NASA and NSF, one of the metrics used to review the efficiency of the organization of a review is the number of days it takes to review a proposal counted from the day it is processed. NSF calls it "dwell time" and the rule in place stipulates that more than 70% of the PIs should be informed of the funding decision within 6 months of the proposal deadline. NSF's reviews routinely hit the 75% mark but as we will show later in this article, this metric is misguided and of limited value. A small and informal survey of NSF PIs shows that most are dissatisfied with the time it takes NSF to take a decision.

With the clock ticking, most reviews have a period of check and compliance during which the proposals submitted are validated and declared compliant to the letter of the program under which they were submitted. This is rather trivial work and involves looking at the number of pages, font, size limit and all sort of formatting limits that are usually put in place for each program. It can also involve some more subtle points, for example, a program which does not allow cost sharing would screen for proposals with cost sharing claims, or some compliance with the NSF requirement that both criteria of "Broader Impact "and "Intellectual Merit" be addressed. This phase is also the moment where proposals should be sorted into panels.

---

[1] Some calls for proposals have rolling deadlines and accept proposals at any time during the year.

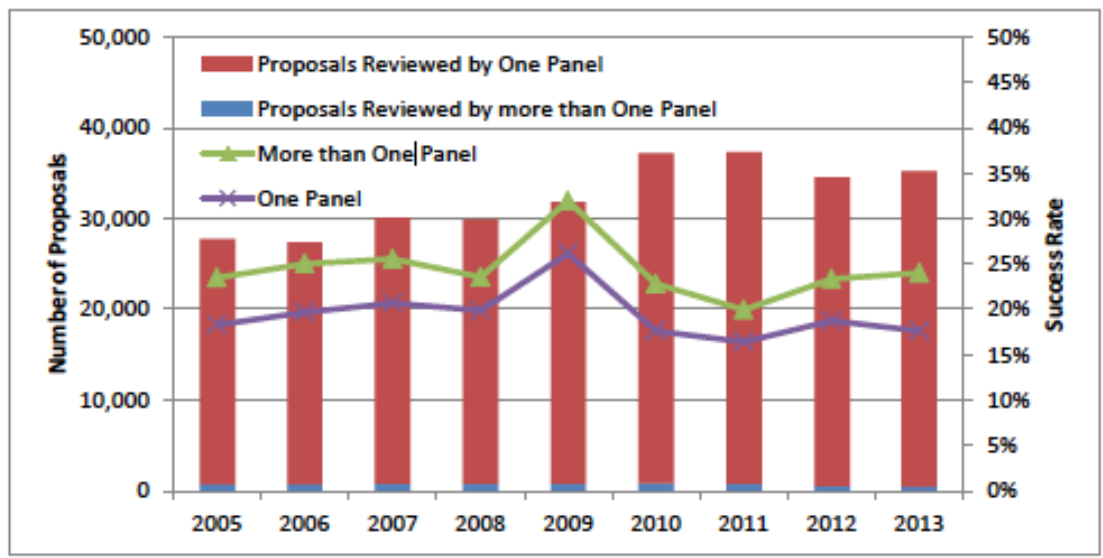

That small step is really the first important part of the process. Its importance varies from program to program but it would be fair to say that a bad "sorting" can doom a proposal just as a "double" sorting, or assigning a proposal to multiple panels for good or arbitrary reasons, can boost the chances of a proposal. The graph above shows that a proposal reviewed in 2 panels had consistently 4 to 6 percentage points higher success rate than proposals reviewed by only one panel[1].

> **Tip #3: (NSF only)**
>
> In light of the 4 to 6 % advantage in the review process when reviewed by two panels, NSF PIs may want to consider asking for their proposals to be reviewed by two panels. PIs should do this only when their request can be justified scientifically and they should write their proposal accordingly.

The number of proposals returned without review for non-compliance is almost negligible. In fact only 35 proposals on a total of around 47,000 proposals submitted were returned without review in FY2013. Considering this statistic, it would be common sense to devote a minimum time to the task. In fact the most important aspect of that stage of any review is, and should be, the sorting of proposals into panels. That is a task that should absorb the time and attention of the organizers.

Sorting the proposals into panels is just one part of the important tasks that need to be done very quickly after the deadline has passed. Most of the time, the sorting is very clear, sometimes it is a bit more involved. All program directors should read more than just the abstracts of the proposals to decide on the assigned panel. The next step is obviously sending the proposals to the panel members leaving them some time to read and research a proposal.

That time should be the minimum period to allow the reviewers assigned to that proposal to familiarize themselves with the work they are asked to judge. Re-assigning a proposal to a panel less than two days before the review, as I have seen it done, all but guarantees that the proposal

---

[1] Reference: National Science Board, FY 2013 Report on the NSF's Merit Review Process, Page 32

will be poorly reviewed. It is true that some reviewers do not read nor submit the reviews for the proposals they are assigned until the very last minute. This does not mean that the organizers should encourage or condone this type of review process.

**Best Practice 1: A proposal should be assigned to a panel at least three weeks before the review. The assignment time should be part of the public report on that proposal.**

A corollary to BP 1 would be that proposals assigned to two panels be carefully discussed, as we have seen that their success rate is significantly larger than proposals reviewed by only one panel. Such decisions should not be taken lightly by fiat of a single program director and it should be extensively justified in writing. This justification should be part of the record linked to that proposal.
Once the proposals have been split into panels, each program director is tasked to find reviewers. There should be no need to explain why the proper reviewers are crucial to the quality of a review. In fact, they are so important that NSF keeps careful data and tab on its reviewers. The grades they give, the reviews they attended and their previous assignments are kept in a large database. NSF tracks reviewers as systematically as they track PI's records.

1-2-2 <u>Forming panels and assigning proposals.</u>

Forming good panels can seem to outsiders like an exercise in "dark magic". It requires a very good knowledge of the field and its community, a good sense of the dynamic of the assembled group and a large dose of patience and resilience in the face of refusals. Conflicts of interests, schedules and simply reluctance from the members of the community to get involved, all play a role in the final composition of panels. The basic guiding principle should be that all PIs submitting their proposals for review have the right to a fair process. This right starts with a panel whose members are qualified to evaluate their proposal(s). **<u>This "qualification" can and should be quantified.</u>**

One method to do so efficiently is to ask the reviewers selected to grade their ability to review the proposals in their panel. The grades vary on a scale ranging from 1 to 5, with a 1 being the reviewer's own field and 5 being a field for which any review, although possible, would require an extensive background and literature search. This is very different from the dubious but popular method of asking panel members to choose the proposals that they would like to review. One obvious reason is that this "self-assignment" system used by any program director, too busy or too lazy, ignores the range of reasons that could make someone volunteer to review a particular proposal. Interest and expertise in the subject matter are not the only reasons why someone would do so and the method is fraught with the dangers of missing an undeclared and undetected conflict of interest. The most important reason to reject such a shortcut that **<u>it does not allow the program director to quantify in any way the value of the reviews.</u>**

The method we are suggesting provides two important data points. For each reviewer, it gives the program director a measure of the "fit" of that reviewer to the panel assigned. For example, a reviewer who grades himself/herself a 4 or 5 for all but one or two proposals should be better used either as an "additional" reviewer only on the two proposals for which he/she is an expert and/or just assigned to a different panel. Most importantly for the fairness of the process, all proposals should have a similar average, ideally less than 2.5, if not at least less than 3. A proposal with an average reviewer rating larger than 3.5 should be flagged and additional reviewers should be recruited specially for that proposal[1].

One trivial objection to this method is that people grade themselves differently but it is our experience that if one defines the grades very clearly as seen in the table below, most reviewers assign their grade consistently. We will also point out that the difference of grading from one reviewer affects all proposals identically.

| | *Suggested grades definition:* |
|---|---|
| 1 | *I am an expert in this field. This is my research. I can provide a competent and objective review of this proposal with minimal background search* |
| 2 | *This is my field although not exactly my research area. I can provide a competent and objective review of this proposal with some background research.* |
| 3 | *This is not exactly my field but I am familiar enough with the subject that I can provide a competent and objective review of this proposal with a reasonable background research.* |
| 4 | *This is not really my field but I think I can provide a competent and objective review of this proposal with extensive background research.* |
| 5 | *This is completely outside my field and expertise. I would be reluctant to provide a review of this proposal even with extensive background research.* |

**Best Practice 2: Adopt a system of self-grading the non-conflicted reviewers on each of the proposals assigned to the panel. The average for each reviewer and average for each proposal should be kept as part of the official review record to prove, should contestation arise afterwards, that all proposals got a fair and equal treatment and that the panel was indeed adequate.**

One last important part of forming panels and **one unfortunately not uniformly in place**, is the complete anonymity of the panel until the review starts. Panelists should not know who else is on the panel until the first day of the review where all the discussions happen with the oversight of an NSF representative. Logistics should be sent to panelists in letters without the names of all the other reviewers to prevent any danger of "cross-talk" before the review.

At NSF where the panels meet under the Federal Advisory Committee Act (FACA) rules, the provision that no-talk should happen without an NSF representative is actually required but not enforced. It also follows that in the NSF system where panelists submit reviews before the actual panel meeting, the default should prevent any panelist to see the other reviews submitted.

**Best Practice 3: Keep the identity of the panelists completely anonymous until the first day of the review. This includes preventing them from reading the reviews submitted by other panelists. This completely eliminates the danger of "pre-review" talks or influence between panelists without the oversight of a review official.**

---

[1] We are not going into the details of the different weights of "external" reviews versus "panelist" reviews. There would be plenty to say about it but it is outside the scope of this article.

1-2-3 <u>Conflicts of interest and their impacts on the review.</u>

Nothing is more important to the good conduct of a review that a strict handling of Conflicts Of Interest (COI)[1]. NSF has very strict rules in places and program directors, both civil servants and rotators, are required to take training that includes a review of these rules. The same is required at NASA with a one-day mandatory training for rotators. Most newcomers do not have any experience or expertise in this area so the training is a great first exposure to the issues. Training material at both NSF and NASA instructs the Program Director to refer as often as necessary to the person in charge of conflicts, the Point of Contact (POC) for COI.

The main difficulty of dealing with COI is the dichotomy between what is legally a conflict and what just "looks bad" but is not legally forbidden. We argue that in the world of reviews, it is better to err on the side of caution and stay clear of everything that could be construed as a conflict. This includes being part of a review in which one has submitted a proposal[2], involves a funding decision over a proposal submitted by one's thesis advisor[3], or having any involvement in a proposal for which one has an official conflict, for example, Program directors assigning reviewers for a proposal with which they are conflicted[4]. This latest is, in our view, the most troublesome because it is based on a erroneous interpretation of an Office of the General Counsel (OGC) document dated December 2003 that states that "a program officer handles a proposal when the officer recommends a decision on it, makes or approves such a decision, or otherwise substantially influences the decision." Because the assignment of reviewers to a proposal is not explicitly listed, assignment of reviewers was deemed not a conflict of interest.[5] This ruling, if indeed it is applied across NSF, could be ground for contests. It is inappropriate for two reasons. The first common-sense reason is that, regardless of the fact that the word "reviewer" appears or not in the OGC guidance, <u>assigning reviewers has a direct impact on the fate of a proposal</u>. It is in fact, one of the reasons NSF is otherwise so careful in its directives to program directors performing this task. "Assigning reviewers" is indeed directly in conflict with the spirit of the OGC directive that deems it illegal to "substantially influence the decision". It is also incredibly ill-advised because assigning reviewers to a proposal that one cannot even read is setting an abysmally low professional standard. Being disqualified from reading the proposal, because of a COI, it is hard to know what type of expertise is needed to provide the needed advice on the value of that proposal. Our recommendation is that if indeed this is the rule across NSF when dealing with conflicts, that it be changed immediately. In summary: A program director in conflict with a proposal should have absolutely nothing to do with it. Only a complete dissociation from the proposal can result in a complaint-proof decision.

> **Tip #4:**
>
> PIs who provide a list of people conflicted might want to reduce the list to less than 5 people (and no large collaborations). PIs do not have to provide justifications but should indicate "personal" or "professional" conflicts. Detailed explanations make an entertaining read but are not required.

---

[1] We do not intend to review in details what constitutes or not a conflict of interest as this is outside the scope of this article.

[2] As a scientist working on astronomy missions, I've been guilty of this in my early days. The practice is still fairly widespread.

[3] Forbidden at NSF but not at NASA.

[4] Forbidden at NASA but not at NSF

[5] Ruling from the NSF's Inspector General office dated February 08, 2015

**Best Practice 4: Any Program Director ruled in conflict with a proposal should have absolutely nothing to do with that proposal. This includes assigning reviewers to that proposal.**

Most readers are familiar with the sight of a conflicted reviewer leaving the discussion's room for a particular proposal. One of the hidden variables of a review is how many people from the panel are left in the discussion after every conflicted reviewer has left the room. This number is not always fixed before the review starts as some conflicts are discovered as the review takes place. It is also different from the number of reviewers assigned to a proposal. Across NSF, and on average, proposals are reviewed (and a written report received and filed) by more than 4 people[1] (it is in most cases only 3 in the AAG review) but these people may or may not be part of the actual discussion when the grading takes place. Any discussion involving the final grading of a proposal should include at least 4 people to diminish the relative weight of a single person's opinion in the voting process.

**Best Practice 5: The number of people participating <u>in the final discussion and grading</u> on a proposal should be tracked and monitored across programs and panels. That number should be made part of the record available on each proposal should a contestation be received after the review.**

1-2-4 Conducting the review – Grading.

As described above, people participating in a review have in principle already read and formed some opinion about the value of the proposals in their panel. The panel review is the time for all reviewers to share and discuss their personal assessment of each proposal. For each proposal, conflicts of interests are announced and dealt wih accordingly. In general, the conflicted reviewer(s) leave(s) the room, the proposal is then discussed for some pre-agreed time and a vote is held to give that proposal a grade. That pre-agreed time depends on a large number of factors: It clearly varies with the number of proposals under consideration and the time assigned to the review but may also be affected by the specifics of the program under which the proposal is submitted and the initial grades received. In some programs for example, reviewers do a pseudo-triage of proposals to avoid spending times on proposals that have no chances of being recommended for funding. A review participant is named the chair and directs the discussions[2]. That different assignation is another difference of roles between the Program Directors and Program Scientists as orchestrating the debates yields considerable more opportunities to shape them. This difference is a direct consequence of the legislation under which each review is organized (see Section 1.2.5).

NSF bases proposal grades and funding recommendations on two distinct criteria. One is the "intellectual merit" of a proposal and the other is its "Broader Impact". Their official definition is given in the official NSF Grant Proposal Guide (GPG)[1]. They are subject to much discussion and interpretation but roughly speaking the intellectual merit of a proposal is its science content. The NSF GPG describes the broader impact as encompassing "the potential to benefit society and contribute to the achievement of specific, desired societal outcomes. " NSF requires that "both criteria are to be given full consideration"[2] but not that both criteria be assigned equal weight in the final grade.

---

[1] This number varies across Directorates and Offices and by review method. Reference: Report to the National Science Board on the National Science Foundation's Merit Review Process Fiscal Year 2012, page 76.

[2] This varies by agencies: At NSF the chair is the Program Director whereas the custom at NASA is to designate a chair among the non-conflicted reviewers.

Reviewers are recruited primarily for their expertise on the science subject (see section 1-2-2 or GPG section III-2-B) and are sometimes not as qualified to judge the scope or validity of a "broader impact" proposal. Anybody who has had to review a wide range of NSF proposals will certainly recognize the difficulty in comparing the "apples and oranges" of the various broader impact proposals in a panel. For most reviewers, the science impact of a proposal is familiar territory, but how can a scientist, recruited for his/her expertise on say "Neutron Stars", compare the "societal impact" of training the only woman graduate student from Nepal in the US to produce a YouTube video?[3]

A panel discussion usually starts with one of the reviewers laying out what the project is, providing both the scientific background and context of the proposal and then describing the team and the exact nature of the work and the resources needed to achieve its goal. As instructed by program directors, the discussion also covers the broader impact of the proposal.

All the other reviewers who have read and graded the proposal prior to the discussion can chime in and extend the description and present their opinion of the proposal. The floor is then opened to all of the other members of the review panel including people who have had no time to read the proposal[4]. Someone is assigned to take notes of the discussion. These notes form the basis of the review report sent after the review to each PI.

The end of the discussion marks a crucial part of the decision process. At that point, a vote is usually taken and a grade assigned. The process here varies from program to program sometimes varying from panel to panel within a single program.
**In essence, the most important part of the decision process is neither quantified nor organized.**

At the discretion of the program directors, there can be no vote at all, but an oral round table with a "must fund", "could fund" or "do not fund" label. There can be integer-only grades varying from 1 to 5 or grades with values allowed in half point increments only or tenth of point increments. The vote can be done orally, Olympic Style, holding a piece of paper, or by secret ballot. Variations exist even within a single program, i.e. the evaluation process of a single proposal depends on the panel. This should never be the case and proposals should be treated consistently and identically, at the very least, across a single program.

---

[1] Reference: "NSF Proposal and Award Policies and Procedures Guide" 2014, Page 72 or GPG, III, A, 1.
[2] Reference: "NSF Proposal and Award Policies and Procedures Guide" 2014, Page 73 or GPG, III, 2.
[3] Those examples are similar to but not actual broader impacts ideas found in proposals.
[4] See Recommendation #1 about the minimum amount of time a panel should have to read a proposal.

**The importance of a secret ballot when it comes to a vote on proposal evaluation cannot be overstated.**

Reviews are done by people connected at least professionally by a shared expertise on a given subject matter. More often than not, reviewers know each other either professionally or personally. They at least know of each other's work. Some reviewers are young professionals trying to make their mark on their field, some are older and more established part of the community. Some reviewers are vocal and very aggressive in their defense or attack of a proposal, some are more subdued. A secret ballot ensures that nobody feels pressured to give a grade higher or lower than they would otherwise just because they fear it would reflect on them. The secrecy of the ballot should be in fact respected until the end of the review when all the averages have been compiled. This ensures a consistency of grading and prevents the "grade inflation" or deflation that can occur as the review progresses and people know what grade is necessary to push or sink a proposal.

**Best Practice 6: The grading of a proposal should be numerical. The grade should be the average of the grades given by the non-conflicted members of the review panel. The vote should be secret (grades written on a piece of paper and passed to the agency representative). The average obtained for each proposal should be kept secret until the end of the first discussion session.** *The range is left at the discretion of the program but we recommend nothing smaller than a 1 to 5 range and using the entire range with at least one decimal place.*

After the first round of voting has taken place and the average point of each proposal is known, there is usually a second round of discussions. While it is very rare to see people completely taken aback by the average-ranked list, it is common for a panel to realize that the merit of a couple of proposals is much higher or lower than what is reflected in the first round of voting. This phase also tries to correct for the "late afternoon" effect when people at the end of a full day of discussions will grade proposals differently from the ones evaluated earlier in the day. Typically, not all proposals are re-examined and discussions of the second round usually concentrate on only a handful of proposals, typically from 20 to 30% of the proposals. This second phase of the panel review is really the finalization of the list that will be given to the Program Director. It is also the phase in which the case for or against a proposal is clarified for the benefit of the person in charge of the report.

1-2-5 Conducting the review – Ranking or no ranking.

Reviewers may not realize it but nothing impacts the way a review is organized more than their own official status. Let's take for example two general programs run and organized by the two main funding agencies for Astronomy, namely ADAP (NASA) and AAG (NSF). At first glance, these are supported by two similar reviews[1]: Proposals are organized in panels, read, evaluated and a list is provided to the Program Director. From that list, the proposals that must, should or could be funded are highlighted. For both programs, the selecting official is not the person in charge of the review but the director of the division or its deputy. For the casual observer, the existing differences if noted at all, seem minor.

---

[1] We are not talking here about the **content** of the two programs but solely on the process.

Nothing could be further from the truth.

All NFS reviews are organized under the Federal Advisory Committee Act (FACA)[1]. The different divisions at NSF do not have the ability to choose the type of reviews they want to organize. They have to operate under the larger umbrella of the entire NSF review process and that means having FACA committees.

According to the General Service Administration (GSA) website, the Federal Advisory Committee Act was enacted in 1972 to "ensure that advice by the various advisory committees formed over the years is objective and accessible to the public." In other words, a committee operating under the regulations spelled out in this act should be open and public. The minutes should be made available to all and so should the composition of the committee. Another consequence of being under "FACA rules" is that the panelists become "civil servants" for the duration of the review.

It would be tempting to dismiss this as a bureaucratic but unimportant point. After all, we know that all NSF review panels are in fact closed and that the minutes are not made public. While it is true that the reviews are closed, per each division's request, reviewer's names are nevertheless made public by law. People interested can look up the entire list of reviewers at: http://facadatabase.gov/default.aspx [2].

On the other side, anybody who has been at a NASA review[3] knows that the instructions given to panelists do not include the task to "rank". In fact, NASA reviewers are instructed **not to rank** the proposals but to grade each according to their **own scientific merits**. Proposals are not to be compared. In other words, and to quote an ex-colleague from NASA, "It is not a defect of proposal A that proposal B is a better proposal". Reviewers are also instructed not to write in the final report that the panel reached a "consensus". This word is banned from any review report and in fact, most program scientists in charge of a review at NASA HQ will go through each and every review summary to make sure that this word indeed does not appear.

From a PI proposer point of view, one of the most visible consequences of these instructions is that it is not at all unusual for a final review from a NASA panel to have many points that seem completely contradictory. I have had my share of phone calls by irate or simply dumbfounded PIs who read their panel summary and cannot understand how, for example, the same part of the proposal appears both as a major strength and a major weakness. The explanation is simply that some members of the panel thought that part was indeed a strong asset of the proposal while some other members in the panel took the exact opposite view; think for example at a common critic of a proposal being "too narrow" versus the other point of view that would find the same approach "extremely well focused". After the panel discussions, both sides agreed to disagree and both points are kept in the panel summary reflecting accurately the panel discussions.

---

[1] This is not true for NASA reviews. There are some NASA's panels that are organized under the FACA rules but none are grant reviews. The list can be found at: http://facadatabase.gov/agency/agencies.aspx and click on NASA.

[2] For example, clicking on "Agencies", "NSF", "Proposals review Panels for Astronomical Sciences", "Committee History", "2014" and "Committee members" will give you the full list of reviewers in 2014 for the Astronomical Sciences Directorate at NSF (all panels and programs)

[3] We are talking here about reviews that are organized and managed directly by HQ. Reviews organized by missions can be very different indeed.

Some reviewers dismiss the precise instructions as an exercise in futility but in fact, it underlines the fundamental point that the review is not under the FACA rule and as such cannot and should not be used to give "advice" to NASA concerning the funding or not of proposals. What the review accomplishes is to give the team of program scientists involved in the process a relatively clear view of the proposals that have scientific merits and the solid backing of their peers. It is the job of the team of program scientists in charge of the review to then formulate the list of funded proposals using their own knowledge of the field, programmatic balance between years, funding availability and all sorts of other factors entering the final decision. The panel grades could be modified to reflect these discussions although most of the time, the grades from the panel are left untouched and only the funding recommendations reflect the program scientists' argument in favor or against a proposal. To be sure, there is a limit to the power of the program scientists. Any proposals initially graded below a grade of around 3.0 or a "Good" in NASA grading parlance, are deemed too weak to warrant funding irrespective of the opinion of the program scientists. Every other proposal could be and sometimes funded.

Let's now look at the way a review changes when it is organized under the FACA rule as they are at NSF. In this case, the panel is officially charged to give advice to NSF. NSF panelists are indeed instructed **to rank proposals** and happily compare proposal A and proposal B. The list provided to NSF program directors is the formal advice given to a funding agency by an official, if very short-lived, advisory committee. This is why NSF reviewers are asked to **sign off the list** at the end of a panel, something they obviously don't do at the end of a NASA panel. That list is part of the record on a panel but it is not communicated back to the proposers. It is worth noting though that there is no reason why this list could not be made public once all the proposals declined have been redacted and otherwise made unidentifiable. The list of funded proposals is already public so the only information that would be added by releasing such list is precisely the transparency of knowing what was the initial advice received by the Foundation on that matter and how it was followed.

> **Tip #5: (NSF only)**
>
> Because NSF's funding is assigned by proposal pressure **only**, PIs in under-represented or under-funded disciplines might want to "coordinate" their proposal's submission in order to reach the "critical mass" needed to have their own panel.

**Best Practice 7: The final list of proposals assigned to each panel should be made public (or at least available upon request) with all information (the entire line) about the proposals not recommended for funding blanked out. Only the proposals pushed forward should appear on the list in the order the panel recommended them and with their overall ranking or grade. Even if most of the list is redacted, it will show the final ranking of funded proposals.**

1-2-6 The final list

Program Directors tend to be very reluctant to deviate from the order of the list and projects tend to be recommended for funding on a basic principle: First the top rated proposals of all the panels linked in a program are pushed forward, then funding the second ranked proposals is made available and so on until the program runs out of money. Of course there can be deviations to this rather simplistic scheme but the ranking or the grade is always the starting point of the discussions on which proposals are recommended for funding.

This reluctance to modify the input of the review panel with a limited knowledge of the general program and programmatic balance across the program, has direct consequences in the conduct of the review. Let's examine the most direct ones.

In the case of NSF Astronomy, proposals are split into 3 large categories by the review panels: Highly Competitive, Competitive and Non Competitive. The first two categories share about 1/3 of proposals, with a share of about 40/60 % in the top and the second category. Proposals in the last category are almost never recommended for funding. These are proposals that are ranked in the lower 2/3 of a panel and with a success rate of less than 15% they just can't compete with proposals ranked higher. It is important to note that there is a **<u>no rule</u>** that prevents a program director to recommend a proposal in this category for funding. This, as the reader may recall, is completely different from the rule in place in NASA stipulating that no Program Officer can overrule a "less-than-good rating" for a proposal. If made, this recommendation would have to be very carefully justified in a report submitted for approval to the selecting official who then can take responsibility for funding such a proposal. This is very rare indeed but can happen.

The top category is where most of the recommended proposals are. There are very concrete consequences of classifying a proposal "Competitive" as opposed to "Highly Competitive". The most obvious one is that it is easier to make the case to recommend a proposal ranked in the Highly Competitive category than in the Competitive category. The more interesting one is that in **some cases** it is actually **more advantageous for a PI** to be ranked at the top of the Competitive category rather than the last of the Highly Competitive one. We will explain this below.

1.2.6.1 <u>EPSCoR: A long and established program.</u>

The NSF Experimental Program to Stimulate Competitive Research (EPSCoR) program was initiated in 1979 to address the glaring disparity in the distribution of Science and Technology research and infrastructure dollars that were being spent by the federal government. The open objective of the program is to help universities and other institutions in states or territories receiving, in the last 3 years, less than 0.75% of the total NSF budget by setting aside a small amount (about 2%) of the research budget. The program turned out to be so successful that several other Federal Agencies, including NASA, have adopted similar but independent programs[1]. In the 35 years of its existence the number of states indentified as "EPSCoR states" has grown from 5 states to 28 and 3 territories (Puerto Rico, Guam and Virgin Islands)[2].

EPSCoR is unique both at NSF and in the world of grant making. We will not go into the details of his inner working. EPSCoR has several programs to achieve the stated goals to increase the amount of federal money that flows to these particular states. One of these strategies, and the one we will talk about here, is called "co-funding". With a budget of about $80 M, the premise of this program is rather simple. Basically "EPSCoR co-funds meritorious proposals that would otherwise not be supported due to lack of availability of funds or other overriding program priorities.[3]". There are two majors points to understand from the co-funding mechanism: 1) It is meant to help funding proposals that are in the "Fund-if-possible" category and not proposals that ranked very highly in the review process and 2) it is designed for proposals that are recommended for funding by the program director in charge of the proposal.

Let's address both points and see how their combination, however good intentioned and logical at first, can lead to non-optimal results.

The EPSCoR rule was designed at a time when the acceptance rate was about 20-25%. In these conditions, the proposals ranked in the "Highly Competitive" or top category in the review could all expect to be recommended for funding. The money was "running out" somewhere near the top of the second category. In these circumstances, the condition that a "proposal needs to be near the top of the second tier" and the requirement that a "proposal would be recommended for funding if money were made available" were overlapping, really almost identical conditions. This is not the case anymore. The money allocation runs dry for proposals inside the first top category (Highly Competitive) but the conditions for being co-funded have not changed. A proposal can only be pushed forward for co-funding if it has been ranked in the "second" category of proposals ("Competitive" in the case of Astronomy).

This makes very clear why in some cases it is <u>better</u> for a proposal to be ranked at the top of the "Competitive" category rather than at the bottom of the "Highly competitive" one. That, of course, only applies to proposals coming from EPSCoR's states but reviewers should be aware of it, as it could manifest itself in a program director insisting that the panel "re-visit" a ranking.

The second condition of the EPSCoR's co-funding seems rather logical. A program director asking EPSCoR for co-funding on a proposal has to commit to fund it, regardless of the outcome of the EPSCoR review There are several deadlines during the year accounting for the different review timescales across NSF's directorates and proposals are reviewed by EPSCoR's office as they are coming in. This condition is of course to ensure that the different program directors only submit to the co-funding program the proposals that they truly believe should be pushed forward but for which they don't have enough money. The problem is that in the current budget environment, the money runs out before ALL the proposals in the highest category are funded. This poses the problem for the program directors on how to access the money available from the EPSCoR's program.

---

[1] Although, these agencies use EPSCoR-like programs, they operate independently (for example, a state can be "EPSCOR" under NSF rules but not under NIH rules). In 1992, the Interagency Coordinating Committee (ICC) was created specifically to improve coordination between these programs.

[2] Reference: EPSCoR 2030. A Report to the National Science Foundation (April 2012).

[3] Reference: http://nsf.gov/od/iia/programs/epscor/EPSCoR_Co-funding_Mechanism.pdf

To explain this problem more clearly let's take the example of two panels A and B. For simplicity, let's assume that both have the same number of proposals and that both reviews resulted in four proposals being identified as Highly Competitive and six others as Competitive. Let's assume that funding is available to support roughly the top three of each panel with some left over[1]. Now let's suppose that the top-ranking proposal for Panel B's "Competitive" category is a proposal submitted from an institution in an EPSCoR's state. The program director for both panels could <u>in principle</u> recommend for funding only the top three proposals of each panel, and submit the proposal from the EPSCoR's state for co-funding. This would require an extensive justification to explain both to the selecting official and the EPSCoR office, why the last proposal of the "Highly Competitive" category of panel B was not recommended for funding when the proposal ranked below is. It is much easier and much less work for the program director to recommend only two proposals in Panel A, four proposals in Panel B and then submit the "Competitive" top proposal for co-funding to the EPSCoR's program. The people in charge of the program have neither access nor knowledge of Panel A and they only see the results of Panel B. In essence, Proposal three from panel A is paying the price of admission for the EPSCoR's proposal and proposal four from panel B benefits from it regardless of its scientific value compared to proposal three from panel A. The irony is, of course, if or when the EPSCoR's office turns down the request for co-funding, as they can and often do, the program director is left with two proposals that are being recommended for funding that would not have been otherwise funded.

We offer in Section 3 our remedy to end all this inappropriate outcome but in the very real possibility that our recommendation on how to finalize a ranked list in a mathematically-derived way won't be followed, our advice to reviewers would be to be very aware of the coded language that you may hear from some program directors. If they would like to be able to ask for co-funding for a proposal coming from an EPSCOR state, they will probably push a bit more to have that proposal ranked at the top of the Competitive category. Program directors will (or at least should) never say openly things like "You should rank this proposal a bit lower" but they may say things like "Have a second look at that list".

Potential PIs should know that EPSCoR's proposals are not the only ones to benefit from a little push or pull. NSF is pretty open with what increases the chances of a proposal to be funded. A non-exhaustive list of what would be valued as good "broader impact" mentions "full participation of women, persons with disabilities, and underrepresented minorities" before "increased public scientific literacy and public engagement with science and technology"[2]. Despite evidence that women's proposals across NSF do on average slightly better than proposals submitted by men[3], reviewers may hear a sentence like "Do you think that we have sufficiently taken into account the different background of the PI" or something like "Please

---

[1] We hope the experts will forgive this simplification. This is just an example to show the mechanism of how things can and do go wrong.

[2] Reference: "NSF Proposal and Award Policies and Procedures Guide" 2014, Page 37 or II-10

[3] Reference: Table 3.1 of the "Report to the National Science Board on the National Science Foundation's Merit Review Process", Fiscal Year 2013, page 56

consider the unconscious bias that you might have had doing the ranking ". All these are basically not too subtle coded-language to tell the review panel "We don't like the order of this list, please try again".

## 2) <u>**Striving for the better method.**</u>

At this point it should be clear to the reader that the step of ranking proposals is a crucial one but one that is subject to many factors that have nothing to do with either their science value or even the collective judgment of a panel. It is also our experience that panel discussions can be fruitless and frustrating especially when they focus on proposals that elicit a strong response from some of the reviewers. A good example would be a proposal that is very poorly ranked by one vocal member of the panel and kind-of-liked by the rest but without a strong voice to defend it.  The discussion usually proceeds as a sorting computer program written by a beginner: Sort proposal A and proposal B, usually by having a discussion and then another vote, then add proposal C and sort Proposal A and Proposal C and if necessary Proposal C and Proposal B. Then the Program director suggests another set of proposals to sort until the entire subset is sorted and/or the panel members are too tired to care. This approach is problematic, not only because it is incredibly slow. The main issue arises because of conflicts of interest. It is very rare that the subset of proposals discussed just happened to be one without any reviewer conflicted. In every "A vs. B" proposal discussion, reviewers conflicted with either the A or B proposal have to exit the room. Of course, they are back when proposal C and D are compared even though the effect of their sorting on the A and B proposals are clear because there is no way to avoid it if they are going to discuss the entire subset.

The other reason this is not an effective method is that the sorting is usually done with the same voting method used in the first stage. **This numerical method has a flaw that even the secret ballot method can't correct: It allows for identical grades**. Nothing in the first round vote prevents a reviewer to give the same grade to several different proposals. At the first stage we argue that this degeneracy is not too crippling because what it is really needed then is a first cut to determine the proposals that are going to be discussed further. Once this list is established though, it becomes crucial that the degeneracy be lifted. In other words, we are asking the reviewers to converge on a list that pleases the majority of them, taking into account the conflicts of interest present in this list.
**Is there a mathematical proven method to do such a thing systematically?  Fortunately, the answer is yes.**

**2.1 The Borda Method**

Jean-Charles de Borda (1733-1799) is the French scientist whose named is attached to a consensus-based voting procedure that does exactly what we describe above[1].
In a Borda count method, people are asked to assign a rank to the list of names or in our case, proposals and PIs. If the list has N names, the top one is credited with N-1 points, the next one receives N-2 and so on until the last one on the list which received 0 points.  **There is no tie**

---

[1] . The method was actually first described by Ramon Llul (1232-1315) but only credited to him after the discovery in 2001 of his manuscripts "*Ars notandi, Ars eleccionis,* and *Alia ars eleccionis*" in which he described what are now known as the Borda count and the Condorcet criterion.

**allowed in a Borda count method and people have to rank the entire list.** The list is then ordered by the total number of points received by each names or proposals on the list. This method has the advantage of producing a ranked ordered list that reflects immediately the consensus of the participants. If the vote is conducted in a secret ballot and it should be as we argue in section 1.2.4, it has also the advantage of reducing the impacts of personality and discussion styles on the final result.

It is easy to see how someone, favored by a slim majority and ranked very poorly by the rest of the panel, would fare very well in a "counting vote" setting. That same person would lose in a Borda Count method where the strength of a candidate resides in its broader appeal. For this, the method is viewed as one freeing any election of the "tyranny of the majority" and striving for a result that is supported by a larger fraction of the reviewers. This is part of the reason why the Borda Count method and some variations are widely used in elections to grant sport awards in the US, both Heisman Trophy and Major League Baseball MVP award used the method, and in many political elections around the world.

Because of the need to take into account potential conflicts of interest, we recommend a Modified Borda Count (MBC) method, in which a reviewer conflicted with a number C of proposals ranks only N-C proposal and his or her top proposal only receives N-C-1 points and the last one receives 0. Note that N is not the total number of proposals in the panel but the smaller subset of proposals that required further discussions. That subset is first identified. For example, one could take the top 1/3 of proposals or whatever subset of proposals is above the "death line" below which no proposal is recommended for funding. The MBC vote is of course done by secret ballot after extensive discussions on the proposals in the subset. **The resulting ranking is final and reflects the consensus** of the majority of people. That is quick, fair, effective and takes into account all the existing conflicts of interest. It is our experience with the MBC method that it is not as effective when the number of proposals exceeds 10. This is probably because of the requirement to break all ties between proposals. In consequence, we don't recommend it for the first stage of a large proposal review.

**Best Practice 8: Implement a unified method for the final ranking of the panel. We suggest a MBC method because it is a proven method combining the advantages of taking care of conflict of interest, ranking and consensus all in one step.**

## 3) Post-review

As we explained at the beginning of this article, the review is only the start of the real work done by program directors. Again, because of the multitude of review types and programs, the detailed review of what happens or should happen after a review would require more than what we intend to say in this article. Let's just briefly mention the guidelines that should be in place to insure that the

> **Tip #6:**
>
> Budget negotiations can be tricky but PIs are rarely served by "selling out" their proposals. Every PI thinks "a little money is better than no money at all" but this is very seldom the actual choice when a PI is asked to change their budget/effort on a proposal. No PI should ever say "I could work with whatever money you can give me". Instead, lay out what the various cuts would do to your project. Also never ask how much money needs to be cut from your proposal's budget to get it through. No good Program Director should/would answer that question. Also remember that a large budget cut (10% at NSF) triggers a mandatory budget resubmission.

post-review is both useful and objective.

As stated previously, at this point in a review, program directors have one or more graded lists of all the proposals submitted. These proposals have been read, discussed and graded and the list can be ranked or not[1]. The program directors are now asked to take this information and give the selecting official a list of the proposals that they are recommending for funding. That final list need not to be identical to the list given by the review panels. In fact, more often than not, it is not. While this difference can open the door to all sorts of abuses, it is, in fact, a healthy guard against taxpayer money waste and, when done professionally, increases the value and return on investment of the review.

We have already seen how the reluctance to deviate from the list can lead to a waste of government money[2]. There are more obvious reasons it is a good thing to allow a relative freedom to program directors in the choices they make in their recommendations. First as mentioned in section 1.2.6, the members of a panel review ignore what other proposals are in the program. Their is a restricted view of the range of proposals submitted so their advice, as objective as it could be, is by definition incomplete.

The panelists have no knowledge of the programmatic balance of a general program and do not know the budget available to the program. For all these reasons, it is crucial that for a sound and defensible selection, the program directors be allowed and even encouraged to use the list from the review as a guide but not a binding document. **Our recommendation (#7) that the initial list be made public or at least available, ensures that the necessary deviations will be extensively documented and justified.**

Program directors at NSF are required to write a document called a "Review Analysis" that is attached to each proposal reviewed. This document contains the reason behind the recommendation and, if recommended, where and when the money will be distributed. A review analysis can range from a boilerplate sentence used if the proposal's rank is within the Non-Competitive category, to several pages if the proposal required extensive negotiations. This review analysis is signed off by the selecting official and represents the official justification for spending public money on that proposal. For this reason, this document should be made public with confidential information redacted when any protest. There should be an equivalent requirement applied to all funding agencies. As we have seen there are many perfectly good reasons why a proposal with a perfect score might not be ultimately funded and why a less than stellar proposal will. All these good reasons should be explained in a document. This will diminish the danger of having abuse by a rogue program scientist. In addition to having a document explaining the recommendation of the program director, we argue that these decisions should be made by a group of at least three program directors, one of them not directly involved in the program but knowledgeable about the science. That would reduce, but of course not eliminate, the risk of abuses while allowing for this crucial liberty of a program director to promote or demote proposals on the ranked list. The review analysis basically then just summarizes the arguments of that mini-committee.

---

[1] See the FACA/non-FACA difference explained in Section 1.2.5
[2] See the example given in section 1.2.6 using the example of a proposal coming from an institution located in an EPCOR state).

**Best Practice 9: All recommendations for funding to the selecting official should be co-signed by more than one program director. We recommend at least three with one not directly connected to the program but knowledgeable about the science. The written document should be made public with confidential information redacted when a protest or inquiry arises.**

It is beyond the scope of this article to explore all the aspects of the entire post-review process but we will mention one aspect of it that should always be in place.

3.1 <u>Review performance measurement</u>

Both NSF and NASA have in place a performance measurement to limit the time it takes them to "act" on proposals. The result is a strict 180-days limit to act upon 70% of the proposals at NSF and a 150-day limit to act on all proposals at NASA. The 70% required at NSF was chosen to allow extra-time for some potential "difficult" proposals. Assembling funding from different sources can, for example, require time and it seemed reasonable to accommodate for that. The Foundation regularly meets this metric[1], even over-performing it, and justifiably regularly issues self-congratulatory reports on that matter. All is well. All except for a small detail: The metric is meaningless.

> **Tip #7:**
>
> While all PIs have the right to protest a panel review, they cannot contest the result but only the process. If the process is indeed flawed, for example a review from an unqualified or conflicted reviewer, then the result may be changed but the first step is to contest the process. Calling to protest the result will accomplish absolutely nothing besides allowing the PI to vent their frustration. Changes of a review result due to a protest are rare but they do occur in about 7% of all cases across NSF. One caveat: Do not contest every proposal review.

With an acceptance rate of less than 15%, it does not take a mathematical genius to see that one could "act upon" 70% of the proposals without actually deciding on any top ranking proposals. As we described in section 1-2-6, proposals falling into the "Do Not Fund" category or the Non Competitive category usually encompass a little less than 70% of the proposals. Accepting the top one or two of a panel simply pushes the number passt that 70% with really none of the hard work done. An informal survey of PIs reveals that most are indeed dissatisfied with the time it takes NSF to get back to them after a proposal deadline. That time is important because it allows PIs to pursue other funding avenues quickly or to start the hiring process for students and other people involved on their proposal.

We suggest another method of judging the efficiency of a review. The first, internal, metric in place should be that of the time between the time proposals are received and a review is organized. This time should not be more than 90 days.
Then the second metric, similar to that in place at NSF now, would impose a limit of about 150 days to act upon more than 85%, or any number around or larger than 100 minus the acceptance rate, of the proposals, and by the 180 day limit all of the proposals have to have been processed. We offer these numbers as first ideas but the following principle should be followed: an internal and short deadline to organize reviews, a longer time to review most of the proposals, an even longer time to finish the "last" troublesome proposals.

---

[1] Except in 2009 when this performance measure was suspended for the last 3 quarters of the year to allow proposals to be funded with the ARRA appropriation.

**Best Practice 10: Implement a performance measurement to include a short internal limit of about 90 days to organize a review. Add two additional longer time ranges to review most and then all proposals. Reviews should be done and closed in at most 6 months.**

## Conclusions & Acknowledgements

We have examined the different aspects of the review process and recommended ten possible ways to improve the process. Some of the recommendations are aimed to make the reviews more efficient and cost-effective. Some are aimed to boost the transparency of the process. Public money for funding research is important and the fact that no unified process exists is something that should be a concern for everyone involved. **Even if none of our recommendations is implemented, we hope that at least the knowledge of what could be improved will trigger healthy discussions in and outside the community. At a time of diminishing grant pool, these discussions are necessary to insure the quality of the research funded.**

The author wishes to thank Dan McKay for his comments on the manuscript. Any remaining typos, errors or French-flavored sentences are my sole responsibility.